\documentclass[prl,twocolumn,floatfix,showpacs,letterpaper]{revtex4}
\usepackage{graphicx}
\usepackage{bm}
\usepackage{threeparttable}
\usepackage{amsfonts}
\usepackage{amssymb}
\usepackage{amsmath}
\usepackage{longtable}
\usepackage{epsfig}

\begin{document}
\topmargin-1.3cm

\title{X-ray absorption signatures of the molecular environment in water and ice}

\author{Wei Chen$^{1}$, Xifan Wu$^{2}$} 
\author{Roberto Car$^{1,2,3}$}
\email{rcar@princeton.edu}
\affiliation{
$^1$Department of Physics, Princeton University, Princeton, NJ 08544, USA\\
$^2$Department of Chemistry, Princeton University, Princeton, NJ 08544, USA\\
$^3$Princeton Institute for the Science and Technology of Materials, Princeton University, Princeton, NJ 08544, USA}
\date{\today}

\begin{abstract}
The x-ray absorption spectra of water and ice are calculated with a many-body approach for electron-hole excitations. The experimental features, including the small effects of temperature change in the liquid, are quantitatively reproduced from molecular configurations generated by ab-initio molecular dynamics. The spectral difference between the solid and the liquid is due to two major short range order effects. One, due to breaking of hydrogen bonds, enhances the pre-edge intensity in the liquid. The other, due to a non-bonded molecular fraction in the first coordination shell, affects the main spectral edge in the conversion of ice to water. This effect may not involve hydrogen bond breaking as shown by experiment in high-density amorphous ice.
\end{abstract}

\pacs{36.20.Ng, 71.15.Pd, 32.10.Dk, 78.20.Ci}

\maketitle
The nature of the hydrogen bond (H-bond) network in water continues to be at the center of scientific debate~\cite{BallNature08}. A few years ago high-resolution x-ray absorption spectra (XAS) probed the local order of liquid and crystalline water phases, including the effect of temperature in the liquid~\cite{Wernet04}. These experiments stirred a storm of controversy because of their interpretation suggesting that a large fraction ($>$80\%) of H-bonds are broken in the liquid~\cite{Wernet04}. This would imply an environment consisting primarily of chains and rings of H-bonds, in stark contrast with the conventional near-tetrahedral picture supported by diffraction~\cite{Head-Gordon06} and other thermodynamic and spectroscopic data~\cite{Eisenberg69,Stillinger80}. The broken H-bond fraction was estimated from the intensity of the pre-edge, which is prominent in the liquid and was believed to be absent in bulk ice. Subsequent x-ray Raman spectra, however, showed that the pre-edge feature is present, albeit with different intensity, not only in the liquid, but also in hexagonal (Ih), cubic (Ic), low-density amorphous (LDA) and high-density amorphous (HDA) ice ~\cite{Tse08}. This experiment reported another interesting observation, namely that water and HDA ice have spectra with the main-edge more prominent than the post-edge, while the opposite behavior occurs in Ih, Ic, and LDA ice. This is puzzling given that both LDA and HDA ice are disordered structures with an insignificant fraction of broken H-bonds. Spectral calculations, based on electronic density-functional theory (DFT) and near-tetrahedral liquid models, correctly predicted the presence of three spectral features in ice and water and associated the enhancement of the pre-edge intensity to broken H-bonds~\cite{Hetenyi04,Prendergast06,Wang06,Jannuzzi08}. However, the agreement with experiment was only semi-quantitative and the calculations did not identify the cause of the significant changes in main- and post-edge spectra. These effects were generically attributed to disorder~\cite{Prendergast06}, but this does not explain why LDA and HDA ice show opposite behavior~\cite{Tse08}. Finally, no attempt was made to discuss the effect of temperature in the liquid spectra, for which contrasting experimental data have been reported~\cite{Wernet04,Smith04,Nilsson05,Smith05}. 
   
In this paper we use liquid structures generated by {\em ab-initio} MD to compute x-ray spectra by adopting the final state rule~\cite{vBarth82} within a many-body formalism for electronic excitations~\cite{Onida02}. 
Our calculations reproduce quantitatively the experimental spectra and their temperature dependence, supporting prior theoretical claims that the spectra are fully consistent with the conventional near-tetrahedral model of the liquid~\cite{Prendergast06}. We associate the x-ray features to molecular excitons strongly influenced by the medium. Broken H-bonds do enhance the pre-edge intensity as pointed out in Ref.~\cite{Wernet04}, 
but the difference between water and ice is also due to other factors commonly associated to 
liquid disorder and does not require an exceedingly large fraction of broken bonds~\cite{Prendergast06}. 
A novel result of our study is the finding that the strong post-edge feature of ice originates 
from a peak in the density of states that disappears in the liquid, as required to explain the corresponding 
spectral changes. This behavior is caused by the presence of a non-bonded fraction within the first molecular 
coordination shell, an effect not directly associated to broken H-bonds, even though the 
short-range order (SRO) of the liquid reflects the partial collapse of the H-bond network. 
We argue that the same effect, in absence of H-bond breaking, should explain the post-edge feature of HDA ice.

Our approach can be viewed as an approximate version of the Bethe-Salpeter equation (BSE) 
for electron-hole excitations~\cite{Onida02}, which was recently applied with success to compute the optical 
absorption spectra of Ih ice~\cite{Hahn05} and liquid water~\cite{Garbuio06}. 
In the present case, we neglect the dynamics of the oxygen 1$s$ core hole, which is localized and long 
lived on the time scale of the absorption process. Then the BSE formalism reduces to a quasi-particle equation for the excited electron in presence of the frozen core hole and the sea of the other electrons~\cite{Rehr04}. 
We solve this equation within the GW approximation, more specifically, the static 
Coulomb-hole and screened exchange (COHSEX) approximation~\cite{Onida02}, which 
is justified for quasi-particle energies in the near edge range of the absorption 
spectrum. In this approach the local exchange correlation potential of DFT is replaced 
by a non-local self-energy operator, which we evaluate by adopting a real space 
representation based on Maximally Localized Wannier functions~\cite{Wu09,support_material}. 

\begin{figure}[htp]
 \begin{center}
  \epsfig{file=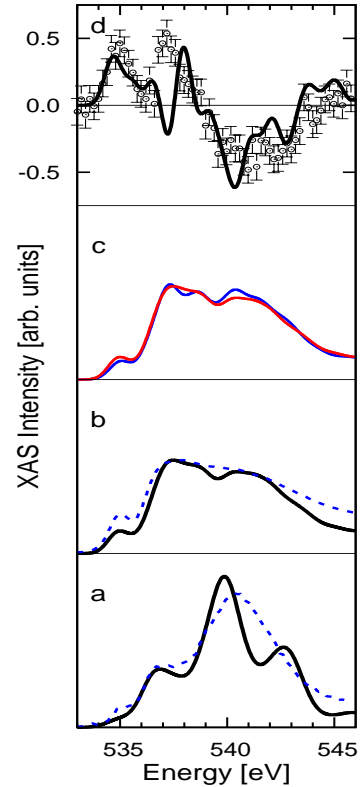,width=2.0in, height = 4.0in} 
 \end{center}
  \caption{ 
Calculated and experimental~\cite{Wernet04} XAS spectra. From bottom to top. (a) Ice: full line (theory, ice Ic), dashed line (experiment, ice Ih, at 95K). (b) Water: full line (theory at 363K), dashed line (experiment at 290K). (c) Theoretical spectra for water at two different temperatures: blue line at 330K, red line at 363K. (d) Difference spectra: theory (full line), experiment (data points with error bars). The temperature difference in the experiment ($\Delta T=65K$) was nearly double that in the calculation ($\Delta T=33K$). Consequently the theoretical data are magnified by a factor of 10 while the experimental data are magnified by a factor of 5 instead than by the factor of 10 used in Ref.~\cite{Wernet04}. A nearly linear dependence on temperature is to be expected for a temperature change small on the energy scale of the structural changes.
}
   \label{xas_cohsex}
\end{figure}

In Fig.~\ref{xas_cohsex} we compare calculated and experimental 
spectra~\cite{Wernet04} of ice and water. The spectra are aligned at the onset and 
all the theoretical curves are multiplied by the scaling factor that adjusts 
the intensity of the main water peak (at T=363K) to the corresponding experimental value (at STP). 
We use uniform Gaussian broadening with a full width at half maximum of 0.6 eV and 0.4 eV, for ice and water, 
respectively, to remove discreteness from the calculated spectra~\cite{footnote_k0} and to 
account for the ignored quasiparticle lifetime and intrinsic broadening effects. 
The liquid spectra are extracted from a single snapshot of 32 molecules at each temperature. We checked with calculations at the DFT level of theory that including more snapshots in the average has minor effect, in agreement with Ref.~\cite{Prendergast06}. At the two temperatures of the simulation the molecules are approximately tetrahedrally coordinated with $\sim7\%$ and $\sim11\%$ of broken H-bonds at T=330 K and T=363 K, respectively, according to the criteria of Ref.~\cite{Luzar96}, or with $\sim14\%$ and $\sim18\%$ broken H-bonds, respectively, according to the criteria of Ref.~\cite{Wernet04}. 

The three main experimental features, pre-edge (533 to 536 eV), near-edge (537 to 539 eV) and post-edge (from 539 eV and beyond), are well reproduced in terms of position, intensity, and spectral width. By contrast, previous DFT calculations, using excited-~\cite{Prendergast06} or full-core hole~\cite{Hetenyi04} potentials,  underestimated the overall spectral width by $\sim$2 eV or more, respectively, with corresponding shifts in the relative peak positions. Somewhat better spectral widths were obtained in half-core hole calculations~\cite{Cavalleri04,Jannuzzi08}, but this approach gave worse agreement in the pre- and main-edge~\cite{Hetenyi04,Wang06}. These inaccuracies are caused by the local DFT exchange-correlation potential that acts equally on all the excited states~\cite{footnote_hybrid}. By contrast the non-local self-energy operator causes higher energy states to experience reduced exchange effects, i.e. a reduced attractive potential, for an overall increase of the spectral width. The agreement between theory and experiment is very good in the liquid, in which we average, at each snapshot, over the 32 local configurations corresponding to the possible oxygen excitations in the supercell. Use of a single high-symmetry local configuration in ice is responsible for the splitting of the post-edge feature into two relatively sharp peaks, at lower and higher energy, respectively, compared to the single, broader, experimental feature~\cite{footnote_better_ice}. In the liquid (Fig.~\ref{xas_cohsex} panels (c) and (d)), a temperature increase enhances pre- and near-edge at the expense of post-edge intensity. Qualitatively, the same effect occurs in the transition from ice to water~\cite{Wernet04}.  Comparison of theoretical and experimental difference spectra in panel (d) shows that the small effects of a temperature change are accurately reproduced, as illustrated by the isosbestic point that falls at 538.4 eV in the calculation, and at 538.8 eV in the experiment. Such a close correspondence in a derivative property would be unlikely barring a good correspondence of the trends in the molecular environment. A different behavior in which {\em both} main- and post-edge features decrease with temperature was reported in nanodroplet experiments~\cite{Smith04}.    

In our calculation, the intensity of the pre-edge feature is weaker than experiment in both water and ice, but the intensity ratio is reasonably reproduced. This reflects approximations as our DFT calculations show that vibrational effects in ice and a larger simulation cell in water would both enhance the pre-edge. An overall underestimate of the pre-edge feature could also result from the assumption of a fully screened core-hole. Moreover the network distortion and the fraction of broken H-bonds in a simulation using classical nuclei and the PBE exchange-correlation functional~\cite{Perdew96} might still be somewhat underestimated in spite of a temperature $\sim$60 K higher than experiment~\cite{Morrone08,support_material}.

\begin{figure}[htp]
\begin{center}
\epsfig{file=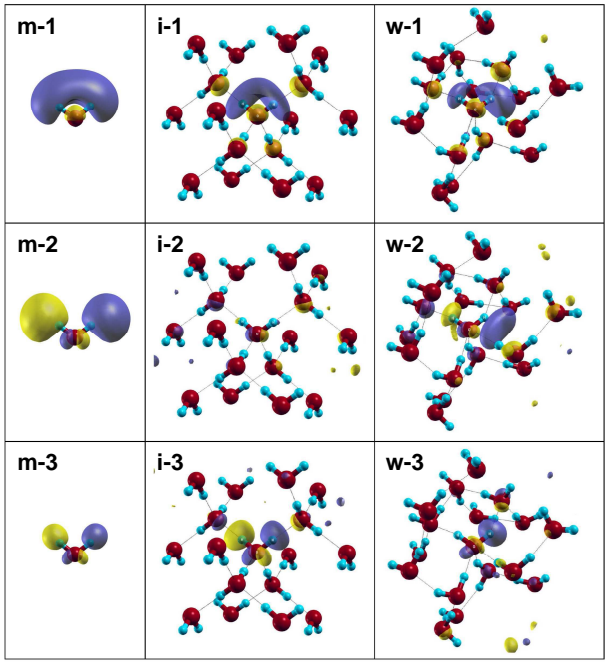, clip, width=0.96\columnwidth}
\end{center}
   \caption{Excited electron wave function in presence of a fully relaxed core hole on the central oxygen atom. The calculation is at the Hartree-Fock level for the monomer (m), at the COHSEX level for ice Ic (i) and water (w). 1: pre-edge, 2: near-edge, 3: post-edge. Only molecules up to the second coordination shell are shown for ice and water. The color difference in the density plots corresponds to a difference in the sign of the wave function.}
   \label{wfc}
\end{figure}

In Fig.~\ref{wfc} we report the COHSEX quasiparticle wave functions of representative pre-edge, near-edge and post-edge states of ice and water. We also report the three lowest unoccupied Hartree-Fock states of the water monomer with a fully relaxed oxygen 1$s$ core hole. The latter are antibonding states of 4$a_1$, 2$b_2$ and 3$b_2$ symmetry, respectively, and correspond to the three lowest energy features of the XAS spectrum in the gas phase~\cite{Pandey06}. The similarity of condensed and gas phase orbitals suggests that the XAS features in ice and water should be attributed to molecular excitons strongly perturbed by the environment. 

We assign the pre-edge to a bound exciton level whose electron orbital is similar to that of the valence exciton that marks the onset of optical transitions in ice~\cite{Hahn05}. Although localized on the excited molecule, this orbital has significant weight also on the four nearest molecules, but the degree of localization may depend on our approximations, such as e.g. fully screened core hole potential and partial neglect of local field effects in the COHSEX calculation. The intact H-bond network of ice enhances the symmetry and reduces the $p$-character of the excited quasi-electron on oxygen, lowering the dipolar transition amplitude and the pre-edge intensity, but the effect is not sufficient to make the feature disappear. This would require the distinction between donor and acceptor bonds to vanish as in ice X, a stable form of ice under extreme pressure~\cite{Hemley87}. From detailed analysis we find that, while on average configurations with broken H-bonds contribute more to the pre-edge intensity than configurations with intact bonds, some distorted tetrahedral configurations can be equally effective. Distortion and fraction of broken H-bonds increase with temperature, enhancing the pre-edge intensity. 

\begin{figure}[ht]
 \begin{center}
  \epsfig{file=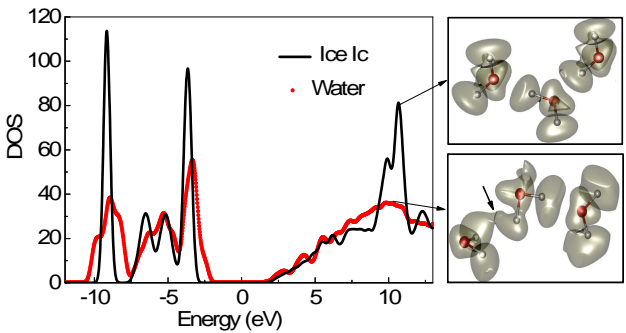,width=0.9\columnwidth} \\
 \end{center}
   \caption{
(color online). Electronic DOS of ice (Ic) and water, calculated with a 4x4x4 $k$-point mesh (ice) and at $k=0$ (water) with uniform Gaussian broadening of 0.3 eV. The conduction band states have positive energy. The insets show a central tagged molecule and two adjacent molecules in ice (top) and water (bottom). Also plotted is the integrated local DOS in correspondence with the DOS peaks. The bottom inset shows overlap, indicated by an arrow, between the antibonding density of the non-bonded molecule on the left and the central molecule, while no overlap occurs between the latter and the bonded molecule at its right. See also Ref.~\cite{support_material}}
   \label{dos}
\end{figure}

The remaining two features in the near- and post-edge, correspond to more delocalized states and originate from several closely spaced levels in our condensed-phase supercell calculations, consistent with the assignment to exciton resonances. While in the monomer and the liquid higher energy states are less localized, the opposite happens in ice, as also reflected in the corresponding relative intensities of near- and post-edge peaks. This phenomenon is not caused by the transition matrix element as both near- and post-edge resonances have strong oxygen $p$-character, but originates from the density of states (DOS). This is illustrated in Fig.~\ref{dos} which depicts the DOS of water and ice (in absence of core hole within DFT). The conduction DOS of ice has a sharp peak at $\sim$8 eV above the band edge corresponding to antibonding molecular orbitals of $b_2$ symmetry. These orbitals have lobes protruding from the covalently bonded hydrogens and on the lone pair side of oxygen. In ice the antibonding state of one molecule does not overlap with the corresponding state of an adjacent molecule, originating a sharp peak in the DOS (Fig.~\ref{dos}). Due to the distinction between antibonding and bonding directions, the overlap between antibonding states of H-bonded molecules is effectively forbidden. By  contrast, when a non-bonded molecule is present in the first molecular coordination shell, overlap occurs, often involving the antibonding H lobes of the two adjacent non-bonded molecules (Fig.~\ref{dos}). A substantial structural change is necessary for that to happen. In the liquid it follows from the partial collapse of the H-bond network. Orbital interaction broadens substantially the peak in the DOS shifting its center of gravity to lower energy, as shown in Fig.~\ref{dos}. As a consequence, post-edge intensity is reduced and oscillator strength is transferred to near-edge. This happens when ice melts and, to a lesser extent, when the temperature is raised in the liquid. Interestingly, the simulations of Ref.~\cite{Prendergast06} found that, when the liquid is quenched to form a glass, substantial post-edge intensity is recovered together with good tetrahedral coordination, albeit in a non-crystalline lattice. This structure should mimic LDA ice, which is obtained experimentally by quenching the liquid. LDA ice has SRO close to cubic and hexagonal ice~\cite{Finney02a}. Not surprisingly, the corresponding x-ray spectra are very similar~\cite{Tse08}. On the other hand, HDA ice, which is obtained by pressure amorphization, sports a main edge similar to the liquid with a pre-edge closer to the crystal~\cite{Tse08}. The latter is consistent with the absence of broken bonds~\cite{Finney02a}. Yet HDA has SRO different from crystalline ice, and actually closer to the liquid, due to the presence of an additional non-bonded molecule, on average, at non-tetrahedral locations within the first coordination shell~\cite{Finney02a}. Arguably this is the reason for the similarity of the main edge of HDA ice and water, which both have molecular density {\em higher} than crystalline ice. In our liquid snapshots more than 10 percent of the molecules have overlapping antibonding orbitals. Interestingly, a sizeable fraction of these molecules (more than 60 percent) have five neighbors. Some of these local configurations correspond to a tetrahedrally bonded central molecule (2 donors, 2 acceptors) with a non-bonded molecule additionally present in the coordination shell. Occasionally, the latter molecule may also be tetrahedrally bonded resulting in a local configuration similar to those of HDA ice~\cite{Finney02a}.

In conclusion, spectral calculations beyond ground-state DFT bear close correspondence to experiment. Our study confirms the sensitivity of the XAS spectra to SRO modifications due to structural transitions and temperature changes. H-bond breaking mainly affects the pre-edge while near- and post-edge are sensitive to an increase of coordination due to a non-bonded molecular fraction. Our work further confirms that the XAS spectrum of water is consistent with the conventional near-tetrahedral picture. The calculated effects of a temperature change in the liquid are illuminating. Theory automatically guarantees the relative normalization of the spectra at different temperatures, a condition that may not be straightforward to enforce in experimental spectra ~\cite{Nilsson05,Smith05}. The excellent agreement of our calculation with the differential spectra of Ref.~\cite{Wernet04} provides independent support of these data. It also suggests that, in spite of the shortcomings of current DFT approximations, the description of the H-bond network provided by {\em ab-initio} MD is fundamentally correct and can capture temperature trends with surprising accuracy. Finally, we note that recent x-ray emission spectra (XES) of liquid water were interpreted as suggesting two coexisting local environments, one low coordinated and the other tetrahedral ~\cite{Tokushima08}. We are unable to directly address this issue because XES cross sections involve core-hole dynamics, which is beyond the present formalism, but we note that water structures from {\em ab-initio} MD do not seem to support a two-liquid hypothesis.

\begin{acknowledgments}
We gratefully acknowledge support from the Department of Energy under grant DE-FG02-05ER46201. We thank G. Galli, A. Nilsson, and L.G.M. Pettersson for useful discussions. 
\end{acknowledgments}

\bibliographystyle{prsty}  
\bibliography{h2o_xas_ref} 

\begin{thebibliography}{10}

\bibitem{BallNature08}
P. Ball, Nature {\bf 452},  291  (2008).

\bibitem{Wernet04}
P. Wernet {\it et~al.}, Science {\bf 304},  995  (2004).

\bibitem{Head-Gordon06}
T. Head-Gordon and M.~E. Johnson, Proc. Nat. Acad. Sci. USA {\bf 103},  7973
  (2006).

\bibitem{Eisenberg69}
D.~S. Eisenberg and W. Kauzmann, {\em The Structure and Properties of Water}
  (Clarendon, Oxford, 1969).

\bibitem{Stillinger80}
F.~H. Stillinger, Science {\bf 209},  451  (1980).

\bibitem{Tse08}
J.~S. Tse {\it et~al.}, Phys. Rev. Lett. {\bf 100},  095502  (2008).

\bibitem{Hetenyi04}
B. Het\'{e}nyi, F.~D. Angelis, P. Giannozzi, and R. Car, J. Chem. Phys. {\bf
  120},  8632  (2004).

\bibitem{Prendergast06}
D. Prendergast and G. Galli, Phys. Rev. Lett. {\bf 96},  215502  (2006).

\bibitem{Wang06}
R.~L.~C. Wang, H.~J. Kreuzer, and M. Grunze, Phys. Chem. Chem. {\bf 8},  4744
  (2006).

\bibitem{Jannuzzi08}
M. Iannuzzi, J. Chem. Phys. {\bf 128},  204506  (2008).

\bibitem{Smith04}
J.~D. Smith {\it et~al.}, Science {\bf 306},  851  (2004).

\bibitem{Nilsson05}
A. Nilsson {\it et~al.}, Science {\bf 308},  793a  (2005).

\bibitem{Smith05}
J.~D. Smith {\it et~al.}, Science {\bf 308},  793b  (2005).

\bibitem{vBarth82}
U. von Barth and G. Grossmann, Phys. Rev. B {\bf 25},  5150  (1982).

\bibitem{Onida02}
G. Onida, L. Reining, and A. Rubio, Rev. Mod. Phys. {\bf 74},  601  (2002).

\bibitem{Hahn05}
P.~H. Hahn {\it et~al.}, Phys. Rev. Lett. {\bf 94},  037404  (2005).

\bibitem{Garbuio06}
V. Garbuio {\it et~al.}, Phys. Rev. Lett. {\bf 97},  137402  (2006).

\bibitem{Rehr04}
J.~J. Rehr, J.~A. Soininen, and E.~L. Shirley, Phys. Scr., T {\bf T115},  207
  (2005).

\bibitem{Wu09}
X. Wu, A. Selloni, and R. Car, Phys. Rev. B {\bf 79},  085102  (2009).

\bibitem{support_material}
{Supporting material on line.}

\bibitem{footnote_k0}
{All calculations are done at the $k=0$ point of the supercell Brillouin Zone.}

\bibitem{Luzar96}
A. Luzar and D. Chandler, Phys. Rev. Lett. {\bf 76},  928  (1996).

\bibitem{Cavalleri04}
M. Cavalleri, M. Odelius, A. Nilsson, and L.~G.~M. Pettersson, J. Chem. Phys.
  {\bf 121},  10065  (2004).

\bibitem{footnote_hybrid}
{The importance of non-local exchange was underlined in a recent cluster
  calculation using a hybrid functional with 52 percent of exact exchange. This
  improved pre- and main-edge, but failed in the post-edge~\cite{Barone08}.}

\bibitem{footnote_better_ice}
{We found, at the DFT level of theory, that vibrational braodening, and better
  $k$-point sampling~\cite{Prendergast06} make the two features merge into a
  single broader peak.}

\bibitem{Perdew96}
J.~P. Perdew, K. Burke, and M. Ernzerhof, Phys. Rev. Lett. {\bf 77},  3865
  (1996).

\bibitem{Morrone08}
J.~A. Morrone and R. Car, Phys. Rev. Lett. {\bf 101},  017801  (2008).

\bibitem{Pandey06}
R.~K. Pandey and S. Mukamel, J. Chem. Phys. {\bf 124},  094106  (2006).

\bibitem{Hemley87}
R.~J. Hemley {\it et~al.}, Nature {\bf 330},  737  (1987).

\bibitem{Finney02a}
J.~L. Finney {\it et~al.}, Phys. Rev. Lett. {\bf 88},  225503  (2002).

\bibitem{Tokushima08}
T. Tokushima {\it et~al.}, Chem. Phys. Lett. {\bf 460},  387  (2008).

\bibitem{Barone08}
G. Brancato, N. Rega, and V. Barone, Phys. Rev. Lett. {\bf 100},  107401
  (2008).

\end{thebibliography}

\end{document}